\documentclass[twocolumn,superscriptaddress,amsmath,amssymb,prl,aps,floatfix]{revtex4-1}
\usepackage{graphicx,color}
\usepackage{epstopdf}
\pdfoutput=1

\begin{document}
%

\title{Influence of an anomalous temperature-dependence of the phase
coherence length on the conductivity of magnetic topological
insulators}

\author{V. Tk\'a\v{c}}
\affiliation{Department of Condensed Matter Physics, Faculty of Mathematics and Physics, Charles University, Ke Karlovu 5, CZ-12116 Prague 2, Czech Republic}
\affiliation{Institute of Physics, P. J. \v{S}af\'{a}rik University, Park Angelinum 9, 040 01 Ko\v{s}ice, Slovak Republic}

\author{K. V\'yborn\'y}
\affiliation{Institute of Physics, Academy of Sciences of the Czech Republic, Na Slovance 2, CZ-18221 Prague 8, Czech Republic}

\author{V. Komanick\'{y}}
\affiliation{Institute of Physics, P. J. \v{S}af\'{a}rik University, Park Angelinum 9, 040 01 Ko\v{s}ice, Slovak Republic}

\author{J. Warmuth}  
\affiliation{Department of Physics, University of Hamburg, D-20355 Hamburg, DE}

\author{M. Michiardi}  
\affiliation{Department of Physics and Astronomy,
Interdisciplinary Nanoscience Center (iNANO),
University of Aarhus, 8000 Aarhus C, DK}

\author{A. S. Ngankeu}  
\affiliation{Department of Physics and Astronomy,
Interdisciplinary Nanoscience Center (iNANO),
University of Aarhus, 8000 Aarhus C, DK}

\author{R. Tarasenko}
\affiliation{Department of Condensed Matter Physics, Faculty of Mathematics and Physics, Charles University, Ke Karlovu 5, CZ-12116 Prague 2, Czech Republic}
\affiliation{Institute of Physics, P. J. \v{S}af\'{a}rik University, Park Angelinum 9, 040 01 Ko\v{s}ice, Slovak Republic}

\author{M. Vali\v{s}ka}
\affiliation{Department of Condensed Matter Physics, Faculty of Mathematics and Physics, Charles University, Ke Karlovu 5, CZ-12116 Prague 2, Czech Republic}

\author{V. Stetsovych}
\affiliation{Institute of Physics, Academy of Sciences of the Czech Republic, Na Slovance 2, CZ-18221 Prague 8, Czech Republic}

\author{K. Carva}
\affiliation{Department of Condensed Matter Physics, Faculty of Mathematics and Physics, Charles University, Ke Karlovu 5, CZ-12116 Prague 2, Czech Republic}

\author{I. Garate}
\affiliation{D\'{e}partement de physique and Institut quantique, Universit\'{e} de Sherbrooke, Sherbrooke (Qu\'{e}bec) CA J1K 2R1}



\author{M. Bianchi}  
\affiliation{Department of Physics and Astronomy,
Interdisciplinary Nanoscience Center (iNANO),
University of Aarhus, 8000 Aarhus C, DK}

\author{J. Wiebe}  
\affiliation{Department of Physics, University of Hamburg, D-20355 Hamburg, DE}

\author{V. Hol\'{y}}
\affiliation{Department of Condensed Matter Physics, Faculty of Mathematics and Physics, Charles University, Ke Karlovu 5, CZ-12116 Prague 2, Czech Republic}

\author{Ph. Hofmann}  
\affiliation{Department of Physics and Astronomy,
Interdisciplinary Nanoscience Center (iNANO),
University of Aarhus, 8000 Aarhus C, DK}

\author{G. Springholz}
\affiliation{Institute of Semiconductor and Solid State Physics, Johannes Kepler University, Altenbergerstrasse 69, A-4040 Linz, Austria}

\author{V. Sechovsk\'{y}}
\affiliation{Department of Condensed Matter Physics, Faculty of Mathematics and Physics, Charles University, Ke Karlovu 5, CZ-12116 Prague 2, Czech Republic}

\author{J. Honolka}
\affiliation{Institute of Physics, Academy of Sciences of the Czech Republic, Na Slovance 2, CZ-18221 Prague 8, Czech Republic}

\date{Apr10, 2018 - edited by KV}


\begin{abstract}
Magnetotransport constitutes a useful probe to understand the interplay  
between electronic band topology and magnetism in spintronics devices    
based on topological materials. A recent theory of Lu and Shen
[Phys. Rev. Lett. 112, 146601 (2014)] on magnetically doped
topological insulators predicts that quantum corrections $\Delta\kappa$ to the  
temperature-dependence of the conductivity can change sign
during the Curie transition. This phenomenon has been attributed to a   
suppression of the Berry phase of the topological surface states at the
Fermi level, caused by a magnetic energy gap. Here, we demonstrate
experimentally that $\Delta\kappa$ can reverse its sign even when the Berry
phase at the Fermi level remains unchanged, provided that the inelastic 
scattering length decreases with temperature below the Curie transition.
\end{abstract}


\maketitle


The material class of topological insulators (TIs) comprises
fascinating fundamental physics and offers potential for spintronic
applications~\cite{Hasan2010,Qi2011,Moore2010}. In three dimensions,
prototype TI materials are bismuth chalcogenides, characterized by a
single gapless topological surface state (TSS).  This TSS, protected
by time-reversal symmetry (TRS), has a Dirac-like linear energy
dispersion and spin-momentum locking~\cite{Lu2011}. Consequently, an electron
travelling around the Fermi circle of a TSS accrues
a quantum (Berry) phase of  $\pi$. In presence of disorder, this Berry phase $\phi_{\text{Be}}$ leads
to an enhancement of the conductivity via quantum interference
(QI). This effect manifests itself most clearly through a negative
magnetoconductance, i.e. weak antilocalization (WAL). In spintronics
applications, it is desirable to open an energy gap $\Delta$ in
the TSS while minimising dopant-induced Coulomb disorder
effects~\cite{Honolka2012}. This can in principle be achieved by doping the TI
with magnetic impurities, provided that the latter order ferromagnetically
in the direction perpendicular to the surface. In such scenario, the
Berry phase is suppressed and the magnetoconductance can change sign
if the Fermi energy lies close to the gap edge, giving rise to weak
localization (WL).

The WAL-to-WL crossover~\cite{HLN1980} has been experimentally
observed in the magnetic field ($B$) dependence of the longitudinal
conductivity $\sigma_{xx}$, for Mn-doped~\cite{Zhang2012}
Bi$_{2}$Se$_{3}$ and other magnetically doped TI thin
films~\cite{Bao2013,Cui-ZuChang2014,Liu2012} in the ferromagnetic regime.
However, a dilemma has emerged concerning the quantum correction to
the zero-field conductivity, $\Delta\sigma_{xx} (T,B=0) = \kappa \ln T$.
While it is expected that WAL-to-WL crossover should be accompanied by
a change from $\kappa<0$ to $\kappa>0$, experiments show
$\kappa>0$ regardless of the concentration of magnetic impurities
(positive $\kappa$ is expected for an ordinary dirty metal under WL
conditions). The theory by Lu and Shen (LS)~\cite{Lu2014} suggests to resolve this dilemma by
taking into account 2D electron-electron interaction (EEI) contributions
$\sigma_{\rm xx}^{\text{ee}}$ with $\kappa^{ee} > 0$, which in
$\kappa=\kappa^{ee}+\kappa^{qi}$ prevail over the QI contribution
$\kappa^{qi}$, but do not override the weak localisation signatures in
$\sigma_{\rm xx}(B)$ (here, the effect of $B$ on $\kappa^{ee}$
can be neglected as Fig. 4(d) in Ref.~\cite{Lu2014} implies). This
theory not only closes the fundamental gap of understanding experimental
data of $\sigma_{\rm xx}(B)$ and $\sigma_{\rm xx}(T)$ but offers a general
procedure to identify the origin of gap $\Delta$ in TSSs. It relies on
field dependent
changes $\Delta\kappa(B) = \kappa(B)-\kappa(0) \propto \kappa^{qi}$
with opposite sign $\kappa^{qi}$ for the WAL and WL phases. 

In this work, we test the LS theory in the case of a magnetically doped
TI (Mn-doped Bi$_2$Se$_3$), and demonstrate that it fails to correctly
identify the reason for gap opening in this particular system and show
what assumption of the LS theory needs to be modified in order to
arrive at the right interpretation. We present field- and
temperature-dependent data of longitudinal ($\sigma_{\rm xx}$) and transverse 
($\sigma_{\rm xy}$) conductances in Mn-doped MBE-grown Bi$_{2}$Se$_{3}$ thin 
films ($x_{\rm Mn} = 0 - 8\% $). The data both above and below the
ferromagnetic ordering temperature $T_{\rm C}$ is analysed in terms of QI
and EEI effects~\cite{Lu2014,Takagaki2012_2} and we observe a distinct
transition from $\Delta\kappa(B=5~T) = +0.5~e^2/\hbar$ at high temperatures to 
$\Delta\kappa(B=5~T) = -0.5~e^2/\hbar$ below $T_{\rm C}$ in apparent 
agreement with predictions of the LS theory concerning the TRS
breaking effect induced by ferromagnetic exchange field
$\langle H_z^{\text{Ex}}\rangle$. This is, however,
not what really happens in Mn-doped Bi$_{2}$Se$_{3}$ as we explain below.
The LS theory as published~\cite{Lu2014} relies on the coherence length
following $l_{\rm \Phi} \propto T^{-p/2}$ with certain temperature-independent
exponent $p$. While this assumption is fulfilled in many systems (in clean
Bi$_{2}$Se$_{3}$, we observe Nyquist-type scaling with $p=1$ for example), our
Mn-doped samples show an anomalous temperature dependence of $l_{\rm \Phi}$.
A maximum appears close to $T=T_{\rm C}$, which is clearly inconsistent 
with $l_{\rm \Phi} \propto T^{-p/2}$ unless different exponents (of opposite 
sign) are taken in the paramagnetic ($T>T_{\rm C}$) and ferromagnetic 
($T<T_{\rm C}$) phases. When this fact is taken into account, the sign
change in $\Delta\kappa$ may be explained without invoking a change in
the Berry phase of the TSS (i.e. with zero or negligible   
$\langle H_z^{\text{Ex}}\rangle$), in agreement with 
other experimental evidence such as ARPES (also explained below).

Mn-doped Bi$_{2}$Se$_{3}$(0001) thin films were grown on BaF$_2$ (111) substrates using MBE with varying Mn concentrations up to the solubility limit of about 8 at. $\%$ ~\cite{Collins-McIntyre2014,Zhang2012} as described in our previous works~\cite{Tarasenko2016}. Film thicknesses $t$ were chosen in the range $t$ = (20 - 40)~nm, just above the thickness limit where top and bottom TSSs of Bi$_{2}$Se$_{3}$ slabs start to directly hybridize and develop a gap $\Delta_{\text{hyb}}$~\cite{Wang2016,DKim2013,Brahlek2014}.

\begin{figure}
\includegraphics[width=85mm]{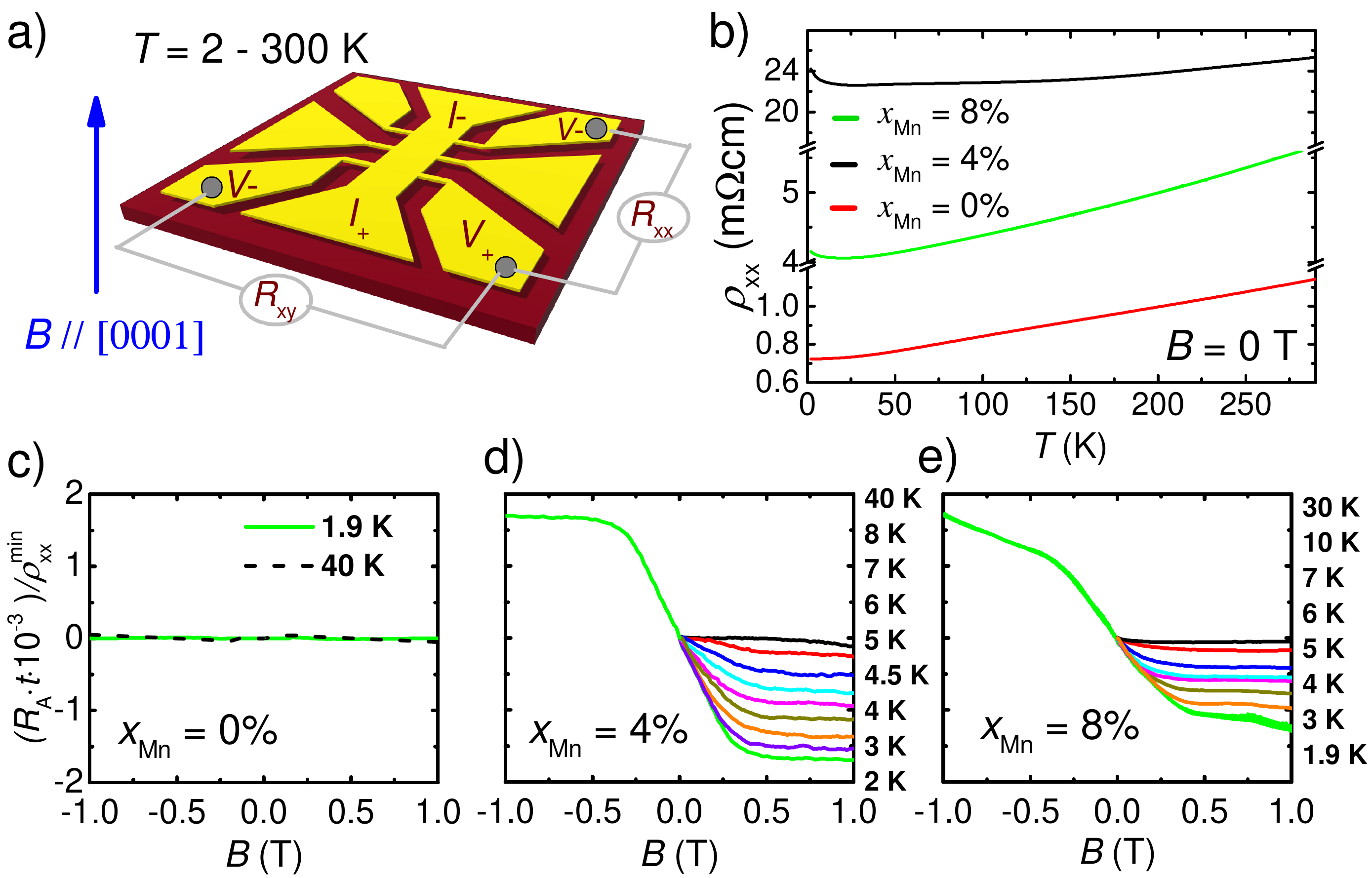} \\
\caption{\label{Fig:Resistivity} (color line) (a) Hall bar measurement geometry of Mn-Bi$_{2}$Se$_{3}$ films. (b) Temperature dependence of the electrical resistivity $\rho_{\rm xx}$ in zero magnetic field for $x_{\rm Mn}$ = 0$\%$, 4$\%$, and 8$\%$ and $t=(40 \pm 5)$~nm. (c)-(e) Respective magnetic field dependence of the anomalous Hall effect $R_{\rm A}$.}
\end{figure}

For magnetotransport measurements we fabricated standard Hall-bars as sketched in Fig.~\ref{Fig:Resistivity}(a) (see also Supporting Information) with longitudinal and transverse voltage probes to determine $\sigma_{\rm xx}$ and $\sigma_{\rm xy}$.
In Fig.~\ref{Fig:Resistivity}(b) we show the temperature dependence of the electrical resistivity $\rho_{\rm xx}=\sigma_{\rm xx}^{-1}$ in zero magnetic field for $t=(40\pm 5)$~nm films with Mn concentrations $x_{\rm Mn}$ = 0$\%$, 4$\%$, and 8$\%$. Resistivities are of the order of 1~m$\Omega$cm, similar to those reported in the quantum transport literature~\cite{Lu2011}. 
Starting at high temperatures we generally observe a resistivity drop with decreasing temperature indicating metallic behavior, while below 30~K all samples exhibit a characteristic increase in resistivity, an effect which we will discuss in the framework of QI and EEIs. 

\begin{figure}
\includegraphics[width=75mm]{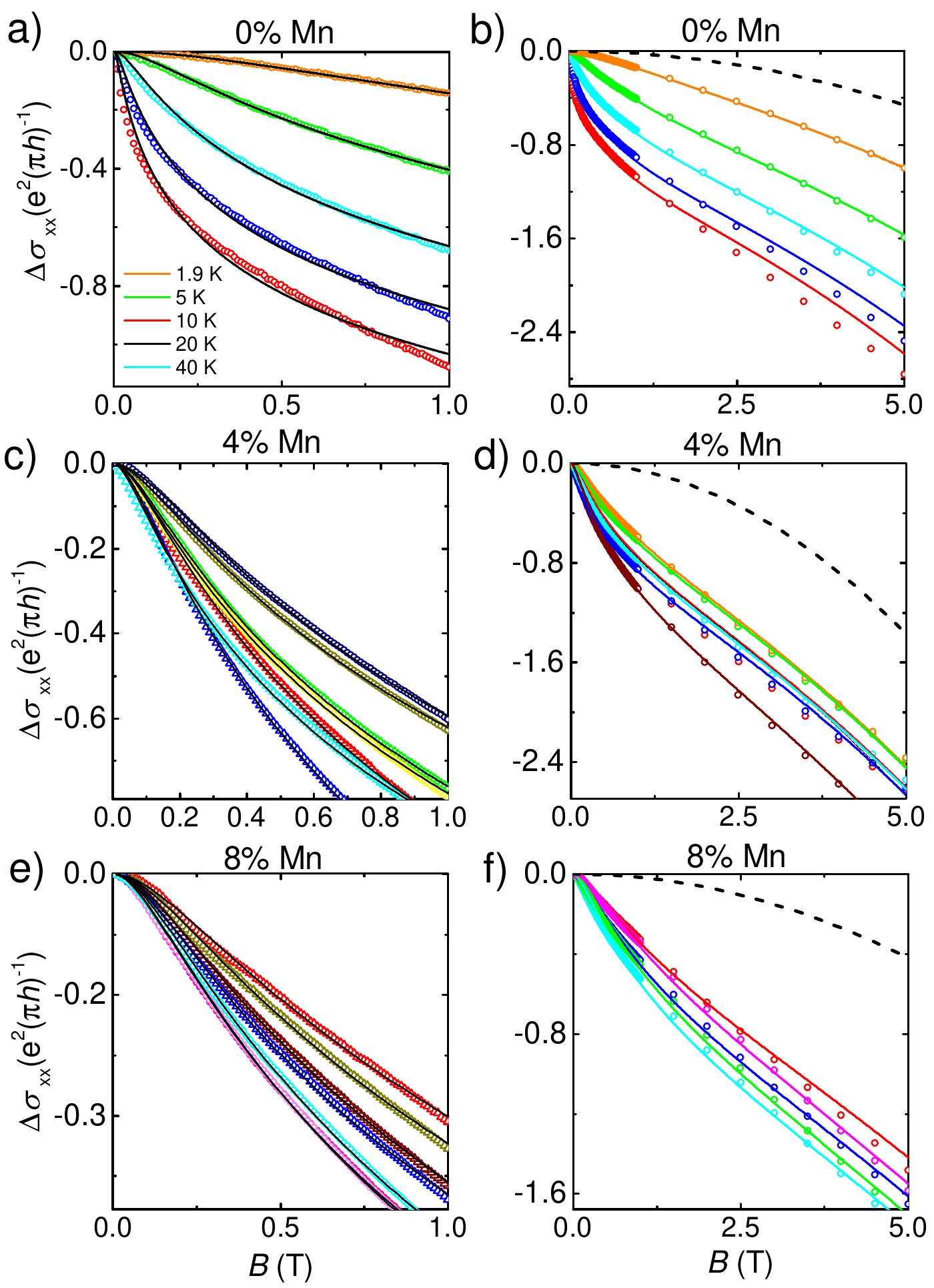} \\
\caption{\label{Fig:Models} (color line) Magnetic field dependence $\sigma_{\rm xx}(B)$ for different temperatures and Mn concentrations (a,b) $x_{\rm Mn}$ = 0 $\%$, (c,d) 4 $\%$, and (e,f) 8 $\%$. The solid lines in (a), (c), and (e) represent fits using Eq.~\ref{eq:three} with $\beta =0$ ($B \le 1$~T), while (b), (d), and (f) leave $\beta$ finite ($B \le 5$~T). $\beta B^{2}$ contributions are plotted separately for $T=2$~K (dashed lines).}
\end{figure}

Charge carrier densities $n$, mobilities $\mu$ as well as magnetic ordering temperatures $T_c$ are derived from Hall measurements in the temperature range between 1.8~K and 40~K. The Hall resistance $R_{\rm xy}$ is given by $R_{\rm xy}(B, T) = R_{\rm N}B + R_{\rm A}(B,T)$, where the linear contribution $R_{\rm N}$ corresponds to the normal Hall effect and $R_{\rm A}$ describes the magnetization dependent anomalous Hall effect (AHE). The results are given in Tab.~\ref{tbl:Hall}. All samples are n-type with carrier concentrations $n=$ 10$^{18}$ - 10$^{19}$ cm$^{-3}$ characteristic for heavily doped Bi$_2$Se$_3$ samples with dominating donor-like Se vacancy defects~\cite{Liu2013,Liu2014_2,Kim2011_2,Cheng2010}.\\
AHE contributions $R_{\rm A}$ in Fig.~\ref{Fig:Resistivity}(c-e)] are shown normalized to $t/\rho_{\rm xx}^{\rm min}$ with $\rho_{\rm xx}^{\rm min}$ listed in Tab.~\ref{tbl:Hall}.
From the onset of the AHE we determine $T_{\rm C}$ = (5$\pm$1)~K and (6$\pm$1)~K for 4$\%$ and 8$\%$, respectively, values close to those found by SQUID for films with $t=500$~nm~\cite{Tarasenko2016}.
\begin{table}
  \caption{Resistivities $\rho_{\rm xx}^{\rm min}$, charge carrier densities $n$, and carrier
    mobilities $\mu$ for different Mn
    concentrations at $T=1.8$~K. The fitting parameter $\beta$ is obtained
    from $\sigma_{\rm xx}(B)$ data using Eq.~\ref{eq:three} at 1.8~K.  }
  \label{tbl:Hall}
  \begin{tabular}{|l|lll|}
    \hline
    & 0 $\%$ Mn & 4 $\%$ Mn & 8 $\%$ Mn\\
    \hline
		$\rho_{\rm xx}^{\rm min}$ [m$\Omega$cm] & 0.72 & 22.58 & 4.06 \\
		\hline
    $n$ [cm$^{-3}$] & 42.7$\times$10$^{18}$ & 1.8$\times$10$^{18}$ & 6.2$\times$10$^{18}$ \\
    $\mu$ [cm$^2$/Vs] & 202.1 & 141.5 & 240.9 \\
        \hline
	$\beta$ [$\mu S/T^2$] & -0.46 & -0.54 & -0.21\\
    \hline
  \end{tabular}
\end{table}
Fig.~\ref{Fig:Models} summarizes the temperature dependence of the magnetoconductance $\Delta \sigma_{\rm xx}(B) = \sigma_{\rm xx}(B) - \sigma_{\rm xx}(0)$ between 1.8~K and 40~K. Independently of Mn concentration, transport properties 
at low fields $B \le 1$~T are dominated by WAL effects $\sigma_{\rm xx}^{\text{qi}}$ leading to cusp-shaped field dependence $\Delta \sigma_{\rm xx}$ with convex curvature. Above 1~T, however, a transition to a concave behavior is observed, typical for Bi$_2$Se$_3$ samples beyond a critical thickness of $\approx$ 20 nm~\cite{Wang2011,Chen2010_2,LiZhang2012}.
For all our samples the cusp-shaped magnetoconductivity peak at low fields is broadened with increasing temperature and finally disappears, which is a consequence of a reduced coherence length $l_{\Phi}$.

For a quantitative estimation of QI effects we rely on an approximation of
the full form Hikami-Larkin-Nagaoka (HLN) theory~\cite{HLN1980}, where additional
spin-orbit and elastic scattering as well as classical cyclotronic magnetotresistance contributions are included as a $\beta B^2$ contribution~\cite{Assaf2013}:

\begin{eqnarray}
\Delta \sigma_{\rm xx} = \frac{\alpha e^2}{2 \pi^2 \hbar}\left[ \psi \left( \frac{1}{2}+\frac{B_{\rm \Phi}}{B} \right) - \text{ln} \left( \frac{B_{\rm \Phi}}{B} \right) \right] + \beta B^2
\label{eq:three}
\end{eqnarray}

where $\alpha \equiv \alpha_{0} + \alpha_{1}$ generally covers both WL
($\alpha_{0}>0$) and WAL ($\alpha_{1}<0$) 2D channel
contributions. $B_{\rm \Phi}$ is the characteristic dephasing field,
$\psi$ the digamma function, and $B\parallel [0001]$ the external
perpendicular magnetic field as shown in
Fig.~\ref{Fig:Resistivity}(a). The dephasing field is defined as
$B_{\rm \Phi} = \hbar/(4el_{\rm \Phi}^{2})$ and $l_{\rm \Phi}$ by
$l_{\rm \Phi} = \sqrt{D \tau_{\rm \Phi}}$, where $D$ is the diffusion
constant and $\tau_{\rm \Phi}$ the phase coherence time. It should be
noted here that by using $\alpha$ as a sole fitting parameter, we
assume $B_{\rm \Phi}$ of WL and WAL contributions ($\alpha_{0}$ and
$\alpha_{1}$) to be the same, which is a common assumption in the
literature~\cite{note2}.

At low fields up to 1~T the simplified form of the HLN theory (case $\beta = 0$) proposed originally by Hikami et al.~\cite{HLN1980} is sufficient to fit the data [see Fig.~\ref{Fig:Models}(a), (c), and (e)]. In the full field range ($B \le 5$~T), however, finite values $\beta<0$ are necessary to account for the concave shape at high fields. Fits in Fig.~\ref{Fig:Models}(b), (d), and (f) for $x_{\rm Mn}$ = 0$\%$, 4$\%$ and 8$\%$, respectively, are generated with values $\beta$ listed in Tab.~\ref{tbl:Hall}. Negative values $\beta$ are expected, e.g. for bulk-like cyclotronic magnetotresistance contributions~\cite{Assaf2013}.  

\begin{figure}
\includegraphics[width=87mm]{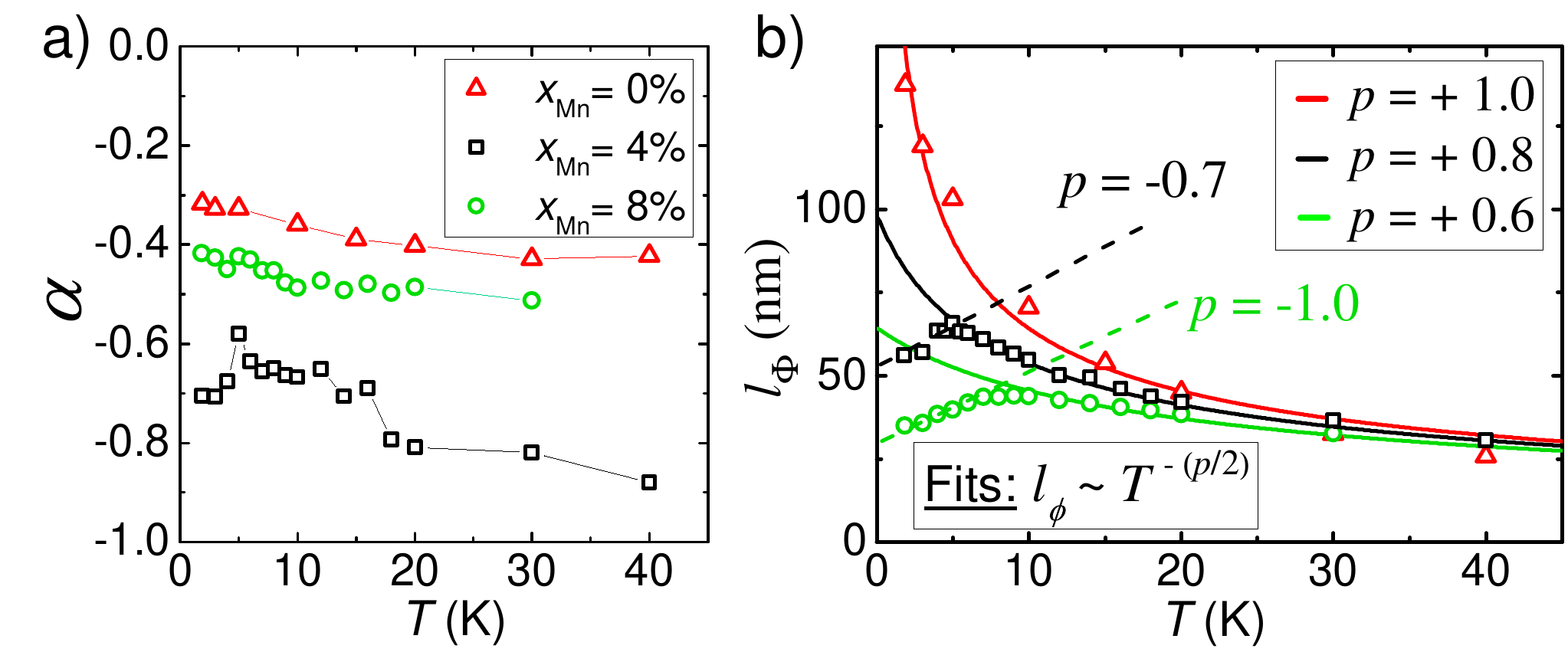} \\
\caption{\label{Fig:Alphas} (color line) Temperature dependence of (a) $\alpha(T)$ and (b) $l_{\Phi}(T)$ for different $x_{\rm Mn}$. Values were derived from the low-field data ($B \le 1$~T) with $\beta =0$. $l_{\Phi}$ is fitted assuming $l_{\rm \Phi} \propto T^{-p/2}$ for $T>T_{\rm C}$ ($p>0$, solid lines) and $T<T_{\rm C}$ ($p<0$, dashed lines).}
\end{figure}
%
We start our discussion with pure Bi$_2$Se$_3$, where at $T=1.8$~K and
$B \le 1$~T ($\beta = 0$) we derive values $\alpha = (-0.32 \pm 0.03)$ and $l_{\Phi}= (137\pm 5)$~nm, which are comparable
to those found in most previous reports~\cite{LiZhang2013,Dey2014}. The
fact that $\alpha<0$ but significantly above $-\frac12$ may be
indicative of the presence of WL contribution from bulk
channels~\cite{Garate2012}. Fits up to highest fields $B \le 5$T
with $\beta = -0.46$ confirm these values within the error margins
(see comparison in Supplementary Table~I) and temperature dependences
(see Supplementary Fig.~S3) show good coincidence. This confirms that
the additional fitting parameter $\beta$ is reasonably independent of
low-field QI contributions $\sigma_{\rm xx}^{\text{qi}}$.  
When increasing the temperature, $l_{\Phi}$ monotonously decreases $\propto T^{-p/2}$ with $p = +(1.0 \pm 0.1)$ (see Fig.~\ref{Fig:Alphas}(a)), as expected for a Nyquist electron-electron scattering mechanism in 2D. 

Modest magnetic doping by Mn at concentrations $x_{\rm Mn}$ = 4$\%$
and 8$\%$ alters QI properties ($\alpha$, $l_{\Phi}$) drastically as
shown in Fig.~\ref{Fig:Alphas}.
We stress here that for magnetic materials with small internal fields
$H^{\text{int}}$, W(A)L effects survive as long as the magnetic length $l_{\rm m}
= \sqrt{\hbar/(e H^{\text{int}})}$ remains larger than the quasiparticle mean
free path, which is the case for diluted magnetic materials such as
(Ga,Mn)As~\cite{Garate2009}.

At finite $x_{\rm Mn}$, $\alpha$ in Fig.~\ref{Fig:Alphas}(b) is generally shifted to more negative values compared to pure Bi$_2$Se$_3$, while
temperature dependences remain weak. 
The corresponding temperature evolution of $l_{\rm \Phi}$ in Fig.~\ref{Fig:Alphas}(a) shows a less steep increase during cooling, which can be described well by lower exponents $p=+(0.8 \pm 0.2)$ and $p=+(0.6 \pm 0.2)$ for $x_{\rm Mn}$ = 4$\%$ and 8$\%$, respectively. It suggests a more efficient spin-dependent dephasing mechanism in the paramagnetic phase compared to a pure Nyquist mechanism for Mn-doped samples as we discuss below. Most interestingly, $l_{\rm \Phi}$ exhibits a characteristic drop when entering the ferromagnetic phase at $T_{\rm C}$ and settles at values $30-70\%$ of those of pure Bi$_2$Se$_3$.
It is important to mention that we checked possible influences of the build-up and saturation of the average perpendicular magnetization $<M_{\rm z}>$ between 0~T and 0.5~T [see AHE signals in Fig.~\ref{Fig:Resistivity}(d,e)] on our fitting results. Fits using HLN theory in the range between 0.5~T and 5~T lead to equivalent results for $\alpha$ and $l_{\rm \Phi}$ within the error margins, which excludes significant influences e.g. of AMR or magnetic domain wall effects on our results.

%
%
%
%

\begin{figure}
\includegraphics[width=86mm]{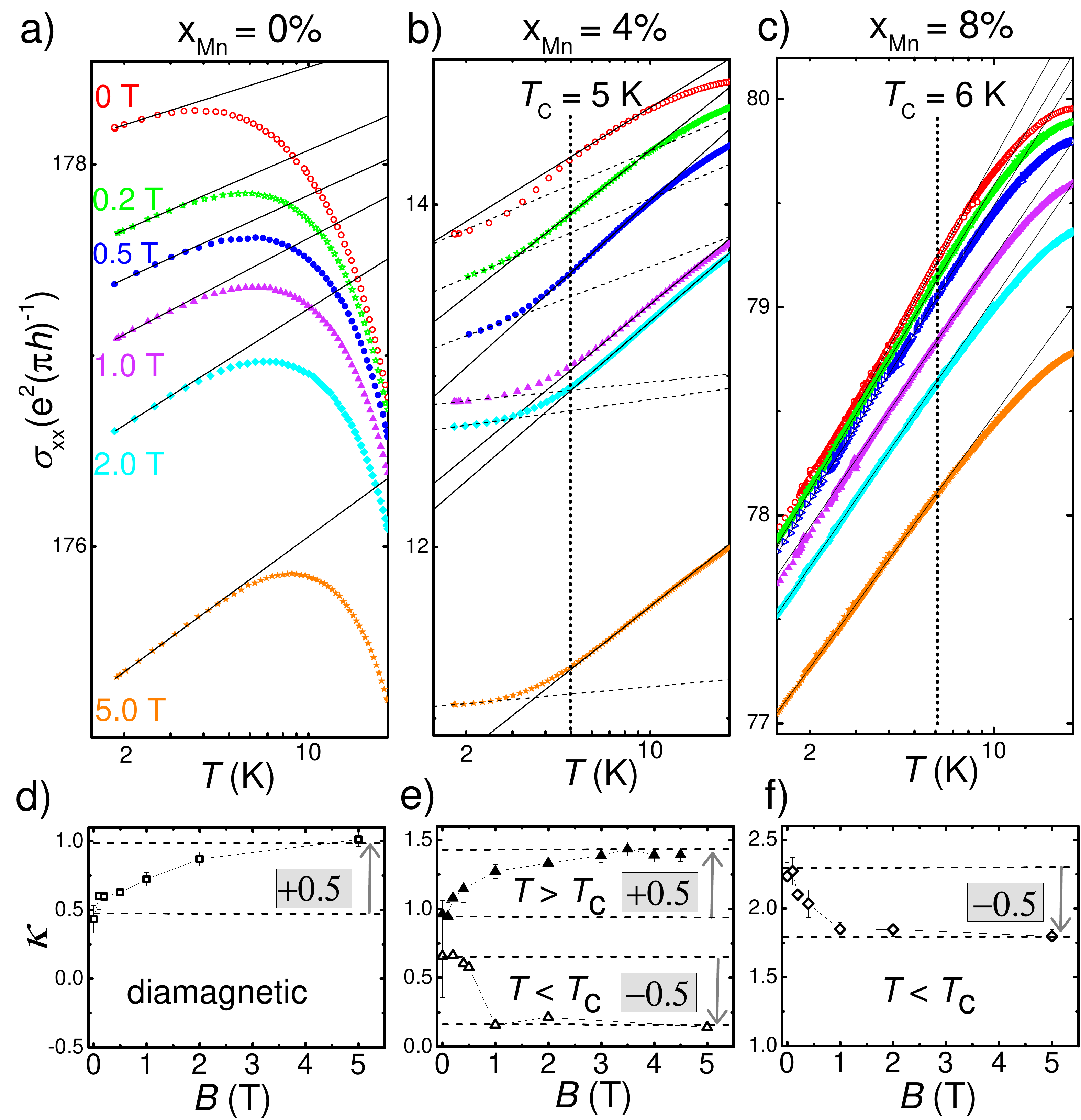} \\
\caption{\label{Fig:LinearFits} (color line) (a-c) Temperature dependence of the electrical conductance in various magnetic fields on a logarithmic scale for $x_{\rm Mn}= $0\%, 4\% and 8\% Mn, respectively. Solid lines are linear fits at $T<T_{\rm E}$. For 4\% and 8\% Mn samples $T_{\rm C}$ is indicated by a dashed vertical line. In (d)-(f) the magnetic field dependence of the slopes $\kappa$ obtained from the linear fits in (a-c) are summarized.}
\end{figure}

In order to test the LS theory, QI properties described by ($\alpha$, $l_{\rm \Phi}$) have to be correlated to the temperature dependence of electrical conductivities $\sigma_{\rm xx}(T, B=const)$ for different constant magnetic fields. Fig.~\ref{Fig:LinearFits} shows the presence of a logarithmic decrease $\sigma_{\rm xx}(T) \propto \text{ln}T$ at lowest temperatures for all samples, reminiscent of a dirty metal regime in 2D. Such a behavior was previously observed in 3D TIs in the absence of the magnetic impurities e.g. for Bi$_{2}$Te$_{3}$~\cite{Liu2014}, Bi$_{2}$Se$_{3}$~\cite{Wang2011} and Sb$_{2}$Te$_{3}$~\cite{Takagaki2012_2}. The logarithmic temperature dependence of the 2D conductivity with EEI corrections can be expressed by~\cite{Lee1985,Altshuler1980}
\begin{eqnarray}
\Delta \sigma_{xx}^{\text{EEI}}(T) = \frac{e^2}{2 \pi^2 \hbar} \sum_{\rm i=0,1} \left( 1-\eta^i_{H}F_i \right) \text{ln} \left( \frac{T}{T_{\rm E}} \right),
\label{eq:EEI}
\end{eqnarray}
where $i$ runs over the number of independent WL and WAL 2D transport channels. $F_i$ is the Coulomb screening factor ($0<F_i<1$),
scaled by a factor $\eta_{H}^{i}$~\cite{Wang2016,Dey2014,Lee1985}. 
$T_{\rm E}$ is the characteristic temperature below which logarithmic EEI corrections dominate $\sigma_{\rm xx}(T)$~\cite{Jing2016},
which is typically in the range of 6 - 10~K~\cite{Jing2016,Takagaki2012} for Bi$_2$Se$_3$, in line with our findings. 
The temperature dependence of $\Delta \sigma_{xx}^{\text{EEI}}$ in Eq.~\ref{eq:EEI} can be quantified by the parameter $\kappa \equiv \left( \pi h / e^2 \right) \partial \sigma_{\rm xx} / \partial $\text {ln}$ T$, and 
according to LS theory, $\kappa$ for each single channel is related to its QI properties $\alpha$ and $p$ via 
\begin{eqnarray}
\kappa = (\alpha\cdot p + 1 - \eta_{H}F)
\label{eq:kappa}
\end{eqnarray}
in the zero magnetic field limit. For $B\gg B_\Phi$, $\kappa \simeq 1-\eta_H F$. Thus, $\Delta\kappa(B)\simeq \alpha p$ at strong fields. 
It is important to note that contributions $\eta_{H}F$ in Eq.~\ref{eq:kappa} are small: for 
gaps $\Delta$ with $\Delta/2E_{\rm F} \le 0.5$ (see discussion of ARPES data below) and permittivities $\epsilon_{\rm r} \approx 100$ one expects $F \le 0.1$, while $\eta_{H}$ remains between 3/4 and 1 for $\Delta$/$2E_{\rm F}$ = 0 and 1, respectively.
EEIs and QIs contribute via $\kappa = \kappa^{ee} + \kappa^{qi}$,
where only the second term significantly depends on the ratios $\Delta/2E_{\rm F}$. For one intact TSS channel with $\alpha = -0.5$ and small screening $F$ one expects $\kappa (B=0) \approx 0.5$ and $\kappa (B=\infty) \approx 1$. For Berry phases $\phi_{\text{Be}}=\pi$, $\kappa$ thus changes by $\Delta \kappa = +0.5$, when $\kappa^{qi}$ contributions are suppressed by the field.

Indeed, for pure Bi$_{2}$Se$_{3}$ we see a perfect match of $\kappa$ with theory predictions for a gapless TSS. As shown in Fig.~\ref{Fig:LinearFits}(a) linear fits are obtained from the logarithmic $\sigma_{\rm xx}(T, B=const)$ data at $T<T_{\rm E}$, and the respective parameters $\kappa(B)$ are plotted as a function of magnetic field [Fig.~\ref{Fig:LinearFits}(d)]. The value in zero magnetic field amounts to $\kappa (B=0) \approx 0.5$, and we see a rise of $\kappa$ with increasing magnetic field according to a suppression of WAL in $\kappa^{qi}$, leading to $\Delta \kappa = +0.5$ and a saturation at about 2-3~T. The result is thus fully in line with Eq.~\ref{eq:kappa} for the experimentally derived 2D Nyquist exponent $p = +1$ for pure Bi$_{2}$Se$_{3}$. 
Saturation fields of $B=3$~T in $\kappa(B)$ agree well with our estimation of $l_{\rm \Phi}$ and resulting dephasing fields $B_{\rm \Phi}$. Our data thus confirms the analysis of magnetoconductance $\sigma_{\rm xx}(B)$, that is the presence of only one single conductive gapless TSS channel~\cite{Lu2014} or - alternatively - two directly or indirectly coupled top and bottom TSSs~\cite{DKim2013,Li2015}.

Fig.~\ref{Fig:LinearFits}(b) and (c) show the respective data at $x_{\rm Mn}$ = 4$\%$ and 8$\%$ for a few fields with according fits for $\kappa(B)$ in the range $T<T_E$ (The full data for all fields are in the supplementary S6). For $x_{\rm Mn}$ = 8$\%$ again a logarithmic temperature dependence $\Delta\kappa \sim \text{ln}T$ in Fig.~\ref{Fig:LinearFits}(c) is visible in the range $T<T_E<T_C$. However, with respect to pure Bi$_2$Se$_3$ the field dependence of $\kappa$ is inverted with $\Delta \kappa$ = -0.5 [see Fig.~\ref{Fig:LinearFits}(f)] as predicted by the theory for a gapped TSS channel. This result at first glance points towards a spectacular evidence for TRS breaking due to ferromagnetic ordering with finite $<M_{\rm z}>$. Our data for $x_{\rm Mn}$ = 4$\%$ with lower $T_C$ but higher $T_{\rm E}$ would support this interpretation; above $T_C$ we find $\Delta \kappa$ $\approx$ +0.5, while below $T_{\rm C}$, $\kappa$ turns into a decreasing behavior $\Delta \kappa$ $\approx$ -0.5.

\begin{figure}
\includegraphics[width=85mm]{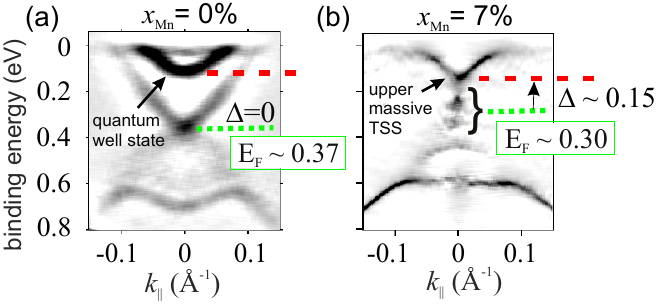}
\caption{\label{Fig:ARPES} (color line) (a) and (b): ARPES for $x_{\rm Mn} =$ 0\% and 7\% measured at $h \nu = 18$~eV and $T=100$~K after 2$^{\text{nd}}$ derivative processing as described in the Supplementary.}
\end{figure}

In the following, we will show that interpreting our data as a
consequence of TRS breaking is tempting but incorrect. Instead, we
attribute the observed sign inversion of $\Delta\kappa$ to a large
change of exponent $p$ around $T_{\rm C}$ (recall that $p$ describes
the scaling of the dephasing time $l_{\Phi}$).
In order to observe a sign inversion in $\Delta\kappa$, TSSs would have to develop a sizeable gap $\Delta$ with $\Delta/2E_{\rm F} \approx 1$. We keep in mind, however, that the predicted size of a purely magnetically induced gap is few meV~\cite{Schmidt2011, Henk2012}. From our surface sensitive ARPES at $T>T_{\rm C}$ in Fig.~\ref{Fig:ARPES} we see strong n-doping with Fermi level shifts $E_{\rm F}$ in the range $0.3-0.4$~eV. It suggests $\Delta/2E_{\rm F} << 1$ unless $\Delta$ would increase to unrealistical values. Thus, from Fig.~\ref{Fig:ARPES}(b) we expect Berry phases $\phi_{\text{Be}}=\pi[1-(\Delta/2E_{\rm F})]$ \cite{Lu2011} to remain close to $\pi$ independent of Mn concentration and magnetic ordering processes. This is in line with our observed prevailing WAL properties with rather constant negative values $\alpha$ above and below $T_{\rm C}$ [Fig.~\ref{Fig:Alphas}(b)]. 
On the other hand, a change of sign in the temperature scaling of the
dephasing time $l_{\Phi}$ as observed in the vicinity of
$T_{\rm C}$ in Fig.~\ref{Fig:Alphas}(a) should indeed lead to an inversion
in $\Delta\kappa$ according to Eq.~\ref{eq:kappa}. Assuming for
simplicity a power law $l_{\Phi}\propto T^{-p/2}$ also below
$T_{\rm C}$, the data can be well described according to a transition
$p=+0.8 \rightarrow -0.7$ ($p=+0.6 \rightarrow -1.0$) for $x_{\rm Mn}$
= 4$\%$ ($x_{\rm Mn}$ = 8$\%$)
[see dashed lines in Fig.~\ref{Fig:Alphas}(a)],
which corresponds closely to the measured inversions
$\Delta\kappa=p\cdot\alpha=\pm0.5$. This interpretation confirms the
validity of the LS theory but shifts the interpretation
of the observed sign inversion of $\Delta\kappa$ from a TRS breaking mechanism
to a temperature-induced change (around $T_{\rm C}$) in phase
decoherent scattering behavior.
Additional proof for the direct correlation between ferromagnetic ordering and transitions $p>0 \rightarrow p<0$ appears when reducing the sample thickness. At $t=20$~nm thickness, ferromagnetism in 4$\%$ Mn samples is suppressed and consequently the paramagnetic QI and EEI behavior with $p = + 0.8$ and $\Delta \kappa$ = +0.5 is extended to lowest temperatures [see Supplementary S3(a)]. We attribute the suppression of ferromagnetism to a weakened bulk-related RKKY coupling in the reduced thickness regime, where $\beta$ values collapse to 0 (Supplementary Table~II) and instead Shubnikov-de Haas oscillations appear as a fingerprint for 2D-dominated transport.

Finally, we comment on microscopic mechanisms that determine the
temperature dependence of the coherence length. In the paramagnetic 
regime $T>T_C$ we find a remarkably good fit to data in
Fig.~\ref{Fig:Alphas}(a) using $l_\Phi=a/\sqrt{T+b_x}$ where $b_x$ is
proportional to $x_{\rm Mn}$ (see Supplementary Fig. S6). Contrary to $l_\Phi\propto T^{-p/2}$
with $p$ as a free fitting parameter, the former temperature
dependence has a clear physical interpretation based on two
uncorrelated mechanisms of inelastic scattering as explained in the
Supplementary information. It suggests that Kondo-type scattering
could be at works, apart from the Nyquist mechanism responsible for
$l_\Phi\propto T^{-1/2}$ observed in pure Bi$_2$Se$_3$ samples.
Such additional scattering mechanism, however, cannot explain the
maximum in $l_\Phi(T)$ around $T_{\text{C}}$. It seems counterintuitive that
perfect ferromagnetic order (or less disorder) should lead to stronger
decoherence and several scenarios such as magnetic field enhanced by
magnetisation or RKKY-type interactions between magnetic
impurities~\cite{Vavilov2003} can be ruled out. We explain these in some
detail in the Supplementary information along with a scattering
mechanism related to spatially inhomogeneous magnetisation within
the sample that gives a larger spin-flip probability at lower
temperatures. Such mechanism would have the potential to explain the
non-monotonous $l_\Phi(T)$ in Fig.~\ref{Fig:Alphas}(a) if it can be
confirmed.  Nevertheless we note that the interpretation of weak
(anti)localisation measurements in the presence of spin-orbit
interaction and magnetic impurities is complicated, in particular when
ferromagnetism is mediated by free carriers~\cite{Garate2009} and a
more detailed study of transport at low temperatures in Mn-doped
Bi$_2$Se$_3$ is likely to reveal interesting physics.

\begin{acknowledgments}

This work was supported by the Czech Science Foundation Grant No. P204/14/30062S. Experiments performed in MLTL (http://mltl.eu/) were supported within the program of Czech Research Infrastructures (Project No. LM2011025). ARPES work was supported by VILLUM FONDEN via the Centre of Excellence for Dirac Materials (Grant No. 11744). VT would like to thank D. Kriegner for the lithography recipe. Scanning electron microscope assistance from K. Uhl\'{i}\v{r}ov\'{a} is gratefully acknowledged. JW, JWa and PH acknowledge funding by the German Research Foundation via the DFG priority programme SPP1666 (grant no. WI 3097/2-2 and HO 5150/1-2). JH acknowledges the Purkyn\v{e} fellowship program. 

\end{acknowledgments}

%
%


\bibliography{Bi2Se3_magnetotransport}

\begin{thebibliography}{39}
\bibitem [{not()}]{note2}%
  \BibitemOpen
  \href@noop {} {}\bibinfo {note} {. The main reason why this assumption is
  commonly made is the complexity of W(A)L fits when $\alpha_0$, $\alpha_1$,
  $B_{\rm \Phi,0}$, and $B_{\rm \Phi,1}$ are left as free
  parameters.}\BibitemShut {Stop}%
\end{thebibliography}

\end{document}